\documentclass[conference]{IEEEtran}
\usepackage[utf8]{inputenc}

\usepackage{xcolor}
\usepackage{color, soul}
\usepackage{enumerate}
\usepackage{dblfloatfix}
\usepackage[shortlabels]{enumitem}
\usepackage{hyperref}
\usepackage{cite}
\usepackage[nomessages]{fp}
\usepackage{tabu}
\usepackage{bm, bbm} 
\usepackage{mathtools}
\usepackage{amsmath, amssymb, amsthm}
\usepackage{nicefrac}
\usepackage{empheq}

\usepackage{booktabs}
\usepackage{graphicx}
\usepackage{multirow}
\usepackage{units}
\usepackage{siunitx}
\usepackage{todonotes}
\usepackage{enumitem}
\usepackage{hyperref}
\usepackage{svg}

\usepackage{alphalph}
\usepackage{etoolbox}
\patchcmd{\subequations}{\alph{equation}}{\alphalph{\value{equation}}}{}{}

\usepackage[noabbrev,capitalize]{cleveref}

\crefname{equation}{}{}
\Crefname{equation}{}{}

\definecolor{myblue}{RGB}{231,144,41}
\newlength{\maxlen}

\makeatletter
\newcommand*\bigcdot{\mathpalette\bigcdot@{.5}}
\newcommand*\bigcdot@[2]{\mathbin{\vcenter{\hbox{\scalebox{#2}{$\m@th#1\bullet$}}}}}
\makeatother

\theoremstyle{definition} % bold title, normal text
\newtheorem{prop}{Proposition}
\theoremstyle{definition} % bold title, normal text

\theoremstyle{remark} % italic title, normal text

\newcommand{\set}[1]{\mathcal{#1}} % for caligraphed set-symbols
\newcommand{\soc}[1]{\left\lVert#1\right\rVert_2} 
\newcommand{\om}{\bm{\omega}}
\newcommand{\Om}{\bm{\Omega}}

\DeclareMathOperator{\Var}{Var}
\DeclareMathOperator{\Sigmart}{\Sigma^{\nicefrac{1}{2}}}

\DeclareMathOperator{\diag}{diag}

\DeclareMathOperator{\Eptn}{\mathbb{E}}
\DeclareMathOperator{\Prb}{\mathbb{P}}

\usepackage{amsmath,amsfonts,amsthm}

\begin{document}
\bstctlcite{IEEE:BSTcontrol} % To activate IEEEtran.bst controls in .bib file

\title{Risk- and Variance-Aware Electricity Pricing
}

% To specify the authors when (number of affiliations > 2)
\author{\IEEEauthorblockN{
Robert Mieth, Jip Kim and Yury Dvorkin}
}

\maketitle

\begin{abstract}
The roll-out of stochastic renewable energy sources (RES) undermines the efficiency of power system and market operations.
This paper proposes an approach to derive electricity prices that internalize RES stochasticity. 
We leverage a chance-constrained AC Optimal Power Flow (CC AC-OPF) model, which is robust against RES uncertainty and is also aware of the resulting variability (variance) of the system state variables.
Using conic duality theory, we derive and analyze energy and balancing reserve prices that internalize the risk of system limit violations and the variance of system state variables.
We compare the  risk- and variance-aware prices on the IEEE 118-node testbed. 
% \vspace{-2mm}
\end{abstract}

\section{Introduction}

Power systems and electricity markets  struggle to accommodate the massive roll-out of renewable energy sources (RES), which are stochastic in nature and impose additional risks on the system operations and market-clearing  decisions. The current industry practice to mitigate these risks is based on procuring additional reserves, which are selected based on exogenous and often ad-hoc policies (e.g., 95-percentile rule in ERCOT, \cite{ercotReserveReq}, or (5+7) rule in CAISO, \cite{nrelwindintegration}). 

Alternatively, such risk assessments can be carried out endogenously, i.e. while optimizing operational and market-clearing decisions, using high-fidelity prediction and historical data parameterizing the RES stochasticity. 
Bienstock et al.~\cite{bienstock2014chance} proposed a \textit{risk-aware} approach to solving an Optimal Power Flow (OPF) problem that uses chance constraints (CC) to internalize the RES stochasticity and risk tolerance of the system operator to violating system constraints.
Since \cite{bienstock2014chance}, the  CC-OPF has been shown to scale efficiently for large  networks \cite{lubin2016robust}, accommodate various assumptions on the RES stochasticity (e.g. parametric distributions and distributional robustness) \cite{lubin2016robust,roald2013analytical,xie2017distributionally}, as well as to accurately account for AC power flow physics, \cite{dvorkin2019chance, roald2016corrective}. 
However, this framework has  primarily been applied to risk-aware operational planning in a vertically integrated environment, neglecting market considerations. 
From a market design perspective, RES stochasticity has been primarily dealt with using  scenario-based stochastic programming, e.g. \cite{morales2012pricing,kazempour2018stochastic,wong2007pricing}, which is more computationally demanding than chance constraints, \cite{bienstock2014chance}. 

With the exception of our recent work in \cite{kuang2018pricing,dvorkin2019chancemarket}, chance constraints have so far been overlooked in electricity pricing applications.
The chance-constrained market design proposed in \cite{dvorkin2019chancemarket} leads to a stable robust equilibrium that, unlike scenario-based approaches in \cite{morales2012pricing,kazempour2018stochastic,wong2007pricing}, guarantees desirable market properties, i.e. welfare maximization, revenue adequacy and cost recovery, under various assumptions on the RES stochasticity. 
Therefore, the resulting energy and reserve prices make it possible to better approximate real-time operating conditions for look-ahead dispatch applications, thus improving consistency between look-ahead and real-time stages. 
However, \cite{dvorkin2019chancemarket} neglects network constraints, an important modeling feature for real-life market applications.

This paper uses a chance-constrained AC OPF (CC AC-OPF) from \cite{dvorkin2019chance} to derive network-aware electricity prices that internalize the RES stochasticity with the intention to produce more accurate signals to market participants. This convex formulation allows the use of duality theory to derive risk-aware marginal-cost-based prices, which are similar to traditional deterministic locational marginal prices (LMPs) based on linear duality, \cite{schweppe2013spot}.  
Furthermore, the CC AC-OPF can explicitly consider reactive power and voltage support services and analyze their role in the deliverabilty of active power, thus supporting the design of a more ``complete'' electricity market, \cite{o2008towards,lipka2016running}.
Completing the market by allowing all assets and services (active and reactive power, reserve capacity, transmission and voltage support) to be transacted, \cite{lipka2016running}, makes it possible to co-align technical needs  and requirements imposed by the physical aspects of power system operations and  price signals received by market participants. 
We also extend the CC AC-OPF to follow a \textit{variance-aware} dispatch paradigm, introduced in \cite{bienstock2018variance}, to compute variance-aware prices and analyze the relationship between the system cost, risk and variance.

\section{Model Formulation}

This paper builds on the AC-CCOPF model presented in \cite{dvorkin2019chance} with model assumptions and modifications explained below. 

\subsection{Preliminaries}

Consider a transmission network with set of nodes $\set{N}$, set of lines $\set{L}$, set of generators $\set{G}$ and set of renewable generators $\set{U}$ (e.g. wind or commercial solar farms).
For simplicity of notation, we assume that each node hosts one conventional and one renewable generator, such that $\set{G} = \set{U} = \set{N}$. 
We denote the set of $PQ$ and $PV$ nodes as $\set{N}^{PQ}, \set{N}^{PV} \subset \set{N}$ and index reference ($\theta V$) node as $i=ref$.
Nodes without generation or with more than one generator can be handled by setting the generation limit to zero or by changing notations, respectively; both modification will not affect the proposed method.
Let vectors $p_G$ indexed as $p_{G,i}$, $p_D$ indexed as $p_{D,i}$, and $p_U$ indexed as $p_{U,i}$, denote the total active power output of conventional generators, the total active power demand and the active power injections from renewable generation at every node. 
The corresponding reactive power injections are denoted $q_G$, $q_D$, $q_U$ and the resulting vectors of net active and reactive power injections are thus given by:
\begin{subequations}%
\begin{align}
& p = p_G - p_D + p_U, \\ 
& q = q_G - q_D + q_U. 
\end{align}
\label{eq:net_injections_det} 
\end{subequations}
In the following, we assume that there is no curtailment of renewable generation and that that all loads $p_D$ are fixed.
We denote $v$ and $\theta$, indexed as $v_i$ and $\theta_i$, as the vectors of voltage magnitudes and voltage angles. 
The range of feasible voltage magnitudes is given as $v \in [v^{min}, v^{max}]$.
Each line in $\set{L}$ is a tuple $ij$ denoting its connected nodes $i,j \in \set{N}$.
For simplicity, we assume a single line between two nodes.
Vectors $f^p$ and $f^q$ indexed as $f^p_{ij}$ and $f^q_{ij}$ denote the active and reactive power flows from node $i$ to node $j$.
Note that $f^p_{ij} \neq f^p_{ji}$ and $f^q_{ij} \neq f^q_{ji}$ due to power losses on the line. 
The vector of apparent power flow limits is denoted as $s^{max}$, indexed by $s_{ij}^{max}$.
We summarize the physical relationship between $p$, $q$, $f^p$, $f^q$, $v$ and $\theta$ as
\begin{align}
    F(p,q,v,\theta) = 0, \label{eq:nonlin_poweflow}
\end{align}
where $F(p,q,v,\theta)$ are the non-linear, non-convex AC power flow equations, \cite[Eq. (2)]{dvorkin2019chance}.

\subsection{Uncertain Power Injections}

We model the real-time deviations from the forecasted renewable active power generation $p_U$ by the random vector $\om$, indexed by $\om_i$, so that the real-time injection from uncertain renewable sources is given by $p_U(\om) = p_U + \om$.
The expected value and covariance matrix of $\om$ are given by $\Eptn[\om] = 0$ and $\Var[\om] = \Sigma$ and we write $\Om = e^{\!\top}\om$ and $S^2 = e^{\!\top}\Sigma e$, where $e$ is the vector of ones. 
The corresponding uncertain reactive power $q_U(\om)$ is linked to the active power generation through a constant power factor $\cos \phi_i$, i.e. $q_{U,i}(\om) = q_{U,i} + \gamma_i \om_i$, where $\gamma_i \coloneqq \nicefrac{\sqrt{1-\cos^2\phi_i}}{\cos\phi_i}$ can either be optimized or fixed in advance. 
Vector $\gamma$ collects all $\gamma_i, i \in \set{U}$.

\subsection{System Response}
\label{ssec:system_response}

To mitigate the effects of $\om$, the controllable generators adjust their output $p_{G}(\om)$ and $q_{G}(\om)$ to maintain the active and reactive power balance.
Subsequently, system state variables $v(\om)$, $\theta(\om)$, $f^p(\om)$, $f^q(\om)$ will respond to those changes based on the system controls and their physical relationship $F(p(\om), q(\om), v(\om), \theta(\om)) = 0$.

As in \cite{bienstock2014chance,dvorkin2019chance,dall2017chance} the response of each generator is given by participation factors $0\leq \alpha_i \leq 1$ that represent the relative amount of the system-wide forecast error ($\Om$) that the generator at node $i$ has to compensate for. 
Therefore, the real-time active power output of each generator is:
\begin{align}
 p_{G,i}(\om) = p_{G,i} - \alpha_i \Om, \label{eq:generator_p_response}
\end{align}
and we require $\sum_{i\in\set{G}} \alpha_i =1$ to balance the system.
Vector $\alpha$ collects all $\alpha_i, i \in \set{G}$. 
The response of reactive power generation $q_{G,i}(\om)$, voltage magnitudes $v_i(\om)$ and voltage angles $\theta_i(\om)$ is determined by the type of node~$i$. 
At $PV$ nodes $v_{i}(\om) = v_{i}, \forall i \in \set{N}^{PV}\!$, is controlled and $q_{G,i}(\om), \theta_i(\om), \forall i \in \set{N}^{PV}\!$, are implicitly determined by power flow equations $F(p,q,v,\theta)$. 
Similarly, at $PQ$ nodes $q_{G,i}(\om) = q_{G,i}, \forall i \in  \set{N}^{PQ}$, is controlled and $v_i(\om), \theta_i(\om), \forall i \in  \set{N}^{PQ}$, are implicitly determined by power flow equations $F(p,q,v,\theta)$.
Finally, at the $\theta V$ node $v_{ref}(\om) = v_{ref}$ and $\theta_{ref}(\om) = 0$. 
Thus, active and reactive power response at the $\theta V$ node is also determined implicitly by power flow equations $F(p,q,v,\theta)$.
The resulting active and reactive power flows are implicitly given by $f^p_{ij}(\om) = f^p_{ij}(v(\om), \theta(\om))$ and $f^q_{ij}(\om) = f^q_{ij}(v(\om), \theta(\om))$.

\subsection{Production Cost}

The production cost of each generator is approximated by a quadratic function, \cite{wood2013power}:
\begin{align}
  c_i(p_{G,i}) = c_{2,i} (p_{G,i})^2 + c_{1,i} p_{G,i} + c_{0,i} \label{eq:quadratic_cost_of_der}
\end{align}
and, for the compactness of derivations, we denote $c_{2,i}=1/2b_i$, $c_{1,i} = a_i/b_i$ and $c_{0,i} = a_i^2/2b_i$.
Given uncertainty $\om$ and the response in \cref{eq:generator_p_response}, the expected operating cost is:
\begin{align}
  \Eptn[ c_i(g_i^P(\om))] = c_i (p_{G,i}) + \frac{\alpha_i^2}{2b_i} S^2. \label{eq:expected_cost}
\end{align}

\subsection{Linearization of AC Power Flow Equations}
\label{ssec:linearization_of_pf_equations}

As discussed in Section~\ref{ssec:system_response}, some system state variables are determined implicitly by the non-linear, non-convex AC power flow equations in \cref{eq:nonlin_poweflow}, which do not permit a direct solution. 
Therefore, we linearize $F(p,q,v,\theta) = 0$ around a given (forecast) operating point using Taylor's theorem as in \cite{dvorkin2019chance}.
Let $(\bar{p}, \bar{q}, \bar{f}^p, \bar{f}^q, \bar{v}, \bar{\theta})$ be the linearization result, then the nodal power injections and line flows are:
\allowdisplaybreaks
\begin{align}
  p_i & = \bar{p}_i + J^{p,v}_{i}(\bar{v}, \bar{\theta}) v + J^{p,\theta}_{i}(\bar{v}, \bar{\theta})\theta \label{eq:p_i_linear}\\
  q_i & = \bar{q}_i + J^{q,v}_{i}(\bar{v}, \bar{\theta}) v + J^{q,\theta}_{i}(\bar{v}, \bar{\theta})\theta \label{eq:q_i_linear}\\
  f_{ij}^p & = \bar{f_{ij}^p} + J_{ij}^{{f^p}, v}(\bar{v}, \bar{\theta}) v + J_{ij}^{fp, \theta}(\bar{v}, \bar{\theta}) \theta \label{eq:fp_ij_linear} \\
  f_{ij}^q & = \bar{f_{ij}^q} + J_{ij}^{{f^q}, v}(\bar{v}, \bar{\theta}) v + J_{ij}^{fq, \theta}(\bar{v}, \bar{\theta}) \theta, \label{eq:fq_ij_linear}
\end{align}
\allowdisplaybreaks[0]%
where $J^{p,v}_{i},J^{p,\theta}_{i},J^{q,v}_{i},J^{q,\theta}_{i},J_{ij}^{{f^p}},J_{ij}^{fp, \theta},J_{ij}^{{f^q}, v}, J_{ij}^{fq, \theta}$ are row-vectors of sensitivity factors describing the change of active and reactive nodal injections as functions of $v$ and $\theta$ derived from the AC power flow linearization.
Similarly, the response of voltages, flows and reactive power outputs to $\om$ can be modeled as (see Appendix \ref{ax:sensitivity_derivation}):
\allowdisplaybreaks
\begin{align}
  q_{G,i}(\om) &= q_{G,i} + [R_i^q(I - \alpha e^{\!\top}) + X_i^q \diag(\gamma)] \om \label{eq:qG_i_sensitivity}\\
  v_i(\om) &= v_i + [R_i^v(I - \alpha e^{\!\top}) + X_i^v \diag(\gamma)] \om \label{eq:v_i_sensitivity}\\
  f^p_{ij}(\om) &= f^p_{ij} + [R_{ij}^{f^p}(I - \alpha e^{\!\top}) + X_{ij}^{f^p} \diag(\gamma)] \om \label{eq:fp_ij_sensitivity}\\
  f^q_{ij}(\om) &= f^q_{ij} + [R_{ij}^{f^q}(I - \alpha e^{\!\top}) + X_{ij}^{f^q} \diag(\gamma)] \om, \label{eq:fq_ij_sensitivity}
\end{align}%
\allowdisplaybreaks[0]%
where row-vectors $R_i^q$, $R_i^v$, $R_{ij}^{f^p}$, $R_{ij}^{f^q}$ map adjustments of the respective variables to active power changes, row-vectors $X_i^q$, $X_i^v$, $X_{ij}^{f^p}$, $X_{ij}^{f^q}$ map adjustments of the respective variables to reactive power changes and $I$ is the identity matrix. 
Note that sensitivity vectors $R_i^q, X_i^q, R_i^v, X_i^v, R_{ij}^{f^p}, X_{ij}^{f^p}, R_{ij}^{f^q}, X_{ij}^{f^q}$ can be zero, if $i$ is a $PV$ or $PQ$ node, and depend on a chosen linearization point.

\subsection{Chance Constrained Optimal Power Flow}

For a given operating point $(p_{G}, q_{G}, v, \theta, \gamma, \alpha)$ the system will respond to any realization of $\om$ according to \cref{eq:generator_p_response}, \cref{eq:p_i_linear,eq:q_i_linear,eq:fp_ij_linear,eq:fq_ij_linear,eq:qG_i_sensitivity,eq:v_i_sensitivity,eq:fp_ij_sensitivity,eq:fq_ij_sensitivity}.
To ensure that this system response does not violate the physical system limits with a high probability, we formulate the following chance constraints:
\begin{align}
  &\Prb(p_{G,i}^{min} \leq p_{G,i}(\om) \leq p_{G,i}^{max}) \geq 1-2\epsilon_p &&i\in \set{G} \label{eq:rawCC_pG_i}\\
  &\Prb(q_{G,i}^{min} \leq q_{G,i}(\om) \leq q_{G,i}^{max}) \geq 1-2\epsilon_q && i\in \set{G} \label{eq:rawCC_qG_i} \\
  &\Prb(v_{i}^{min} \leq v_{i}(\om) \leq v_{i}^{max}) \geq 1-2\epsilon_v, && i\in \set{N} \label{eq:rawCC_v_i}\\
  & \Prb((f_{ij}^p(\om))^2 + (f_{ij}^q(\om))^2 \leq (s_{ij}^{max})^2) \geq 1-\epsilon_f \quad \hspace{-1cm}&&\hspace{1cm} ij \in \set{L}, \label{eq:rawCC_flow}
\end{align}
where $\epsilon_p$, $\epsilon_q$, $\epsilon_v$, $\epsilon_f < \nicefrac{1}{2}$ can be chosen to tune the risk level associated with the individual chance constraints.
Using \cref{eq:qG_i_sensitivity,eq:v_i_sensitivity,eq:fp_ij_sensitivity,eq:fq_ij_sensitivity}, we can obtain computationally tractable reformulations of chance constraints \cref{eq:rawCC_pG_i,eq:rawCC_qG_i,eq:rawCC_v_i,eq:rawCC_flow}, \cite{dvorkin2019chance,dall2017chance,bienstock2014chance}, and formulate the deterministic equivalent of the CC AC-OPF:
\allowdisplaybreaks
\begin{subequations}
\begin{align}
  & \text{EQV-CC}: &&\min_{\substack{p_{G}, q_{G} \\ v, \alpha, \theta }} \sum_{i\in \set{G}} c_i (p_{G,i}) + \sum_{i\in \set{G}} \frac{\alpha_i^2}{2b_i} S^2 \label{det_accc:objecive}
\end{align}
% \vspace{-0.8cm}
\begin{align}
& \text{s.t.} && \nonumber \\
% Power Balances
  & (\lambda^p_i,\lambda^q_i): \hspace{-0.5cm}&&\hspace{0.5cm} \text{\cref{eq:p_i_linear}, \cref{eq:q_i_linear}} \label{det_accc:lambda} \\
% Flow expressions
  & (\beta^p_{ij},\beta^q_{ij}): 
    \hspace{-0.5cm}&&\hspace{0.5cm} \text{\cref{eq:fp_ij_linear}, \cref{eq:fq_ij_linear}} \label{det_accc:beta} \\
% Balancing Adequacy
   & (\chi): &&
    \sum_{i \in \set{G}} \alpha_i = 1 \label{det_accc:chi}\\
% Active power at non-slack generators
  & (\delta_i^{p,+}): &&
    p_{G,i} + \alpha_i z_{\epsilon_p} S \leq p_{G,i}^{max} && i \in \set{G} \label{det_accc:delta_p_plus_i} \\
  & (\delta_i^{p,-}): &&
    -p_{G,i} + \alpha_i z_{\epsilon_p} S \leq -p_{G,i}^{min} && i \in \set{G} \label{det_accc:delta_p_minus_i} \\ 
% Reactive power at all generators (Only Reactive power at PV nodes is uncertain)
  & (\delta_i^{q,+}): &&
    q_{G,i} + z_{\epsilon_q} t_i^{q} \leq q_{G,i}^{max} && i \in \set{G} \label{det_accc:delta_q_plus_i} \\ 
  & (\delta_i^{q,-}): &&
    -q_{G,i} + z_{\epsilon_q} t_i^{q} \leq -q_{G,i}^{min} && i \in \set{G} \label{det_accc:delta_q_minus_i} \\ 
  & (\zeta_i^q): &&
    \soc{(R_i^q\!-\!\rho_i^q e^{\!\top}\!\!+X_i^q \diag(\gamma)) \Sigmart}\!\leq t_i^{q} 
    \hspace{-2cm}&&\hspace{1.7cm} i \in \set{G} \label{det_accc:zeta_i_q} \\
  & (\nu_i^q): && R_i^q \alpha = \rho_i^q && i \in \set{G} \label{det_accc:nu_i_q} \\
% Voltages 
  & (\mu_i^{+}): &&
    v_{i} + z_{\epsilon_v} t_i^{v} \leq v_{i}^{max} && i \in \set{N} \label{det_accc:mu_i_plus}\\
  & (\mu_i^{-}): &&
    -v_{i} + z_{\epsilon_v} t_i^{v} \leq -v_{i}^{min} && i \in \set{N} \label{det_accc:mu_i_minus}\\ 
  & (\zeta_i^v): &&
    \soc{(R_i^v \!-\! \rho_i^v e^{\!\top}\!\! + X_i^v \diag(\gamma)) \Sigmart}\! \leq t_i^{v} 
    \hspace{-2cm}&&\hspace{1.7cm} i \in \set{G} \label{det_accc:zeta_i_v} \\
  & (\nu_i^v): && R_i^v \alpha= \rho_i^v && i \in \set{N}\label{det_accc:nu_i_v} \\
% Line Flows
  & (\eta_{ij}): &&
    (a_{ij}^{f^p})^2 + (a_{ij}^{f^q})^2 \leq (s_{ij}^{max})^2, && ij \in \set{L} \\
  & (\xi^{{f^p},+}_{ij}): &&
     -a^{f^p}_{ij} + z_{\nicefrac{\epsilon_f}{2.5}} t^{f^p}_{ij} \leq f^p_{ij} && ij \in \set{L} \label{det_accc:xi_ij_p_plus}\\ 
  & (\xi^{{f^p},-}_{ij}): &&
    -a^{f^p}_{ij} + z_{\nicefrac{\epsilon_f}{2.5}} t^{f^p}_{ij} \leq -f^p_{ij} && ij \in \set{L}  \\
  & (\xi^{{f^p},0}_{ij}): &&
    z_{\nicefrac{\epsilon}{5}} t^{f^p}_{ij} \leq a^{f^p}_{ij} && ij \in \set{L} \\
  & (\xi^{{f^q},+}_{ij}): &&
    -a^{f^q}_{ij} + z_{\nicefrac{\epsilon_f}{2.5}} t^{f^q}_{ij} \leq f^q_{ij}, && ij \in \set{L} \\
  & (\xi^{{f^q},-}_{ij}): &&
    -a^{f^q}_{ij} + z_{\nicefrac{\epsilon_f}{2.5}} t^{f^q}_{ij} \leq -f^q_{ij}, && ij \in \set{L} \\
  & (\xi^{{f^q},0}_{ij}): &&
    z_{\nicefrac{\epsilon_f}{5}} t^{f^q}_{ij} \leq a^{f^q}_{ij} && ij \in \set{L} \label{det_accc:xi_ij_q_0}\\
  & (\zeta_{ij}^{\diamond}): &&
    \soc{(R_i^\diamond - \rho_i^\diamond e^{\!\top} + X_i^\diamond \diag(\gamma)) \Sigmart} \leq t_i^{\diamond} 
    \hspace{-2cm}&&\hspace{2cm} \nonumber \\
    & && &&\hspace{-1.2cm} ij \in \set{L}, \diamond = f^p, f^q \label{det_accc:zeta_ij_f} \\
  & (\nu_{ij}^{\diamond}): && R_{ij}^\diamond \alpha= \rho_{ij}^\diamond && \hspace{-1.3cm}ij \in \set{L}, \diamond = f^p, f^q, \label{det_accc:nu_ij_f} 
\end{align}%
\label{mod:main_acccopf}%
\end{subequations}%
\allowdisplaybreaks[0]%
where Greek letters in parentheses in \eqref{det_accc:lambda}--\eqref{det_accc:nu_ij_f} denote dual multipliers of constraints. 
Objective \cref{det_accc:objecive} minimizes the expected cost as in \cref{eq:expected_cost}.
Eqs. \cref{det_accc:lambda,det_accc:beta} are the active and reactive power balances and flows based on the linearized AC power flow equations.
Eq. \cref{det_accc:chi} is the balancing reserve adequacy constraint and \cref{det_accc:delta_p_plus_i}--\cref{det_accc:nu_ij_f} are the deterministic reformulation of chance constraints \cref{eq:rawCC_pG_i,eq:rawCC_qG_i,eq:rawCC_v_i,eq:rawCC_flow}, \cite{dvorkin2019chance}.
Constraints \cref{det_accc:delta_p_plus_i,det_accc:delta_p_minus_i} limit the active power production $p_{G,i}$ and the amount of reserve $\alpha_i z_{\epsilon_p} S$ provided by each generator, \cite{dvorkin2019chancemarket,mieth2019distribution}.
Risk parameters are given by $z_{\epsilon} = \Phi^{-1}(1-\epsilon)$, where $\Phi^{-1}(1-\epsilon)$ is the $(1-\epsilon)$-quantile of the standard normal distribution, if $\om$ follows a normal distribution.
Although less restrictive assumptions on the distribution of $\om$ can be invoked in \eqref{mod:main_acccopf}, e.g. by means of non-Gaussian parametric distributions \cite{roald2013analytical} or distributionally robust formulations \cite{dvorkin2019chancemarket,lubin2016robust}, this paper assumes  normally distributed forecast errors  for the sake of presentation clarity.
The standard deviation of reactive power outputs, voltage levels and flows resulting from the uncertainty and the system response is given by the SOC constraints \cref{det_accc:zeta_i_q,det_accc:zeta_i_v,det_accc:zeta_ij_f}. 
Given the convexity of the SOC constraints, auxiliary variables $t_i^q$, $t_i^v$, $t_{ij}^{f^p}$, $t_{ij}^{f^q}$ relate these standard deviations to the reactive output limits \cref{det_accc:delta_q_plus_i,det_accc:delta_q_minus_i}, voltage bounds \cref{det_accc:mu_i_plus,det_accc:mu_i_minus} and flow limits \cref{det_accc:xi_ij_p_plus}--\cref{det_accc:xi_ij_q_0}.
Due to its quadratic dependency on the uncertain variable, the chance constraint in \cref{eq:rawCC_flow} requires a more complex reformulation than \cref{eq:rawCC_pG_i,eq:rawCC_qG_i,eq:rawCC_v_i}.
To accommodate this reformulation, we follow \cite{dvorkin2019chance} and introduce auxiliary variables $a_{ij}^{f^p}$, $a_{ij}^{f^q}$ and risk parameters $\nicefrac{\epsilon_f}{2.5}$ and $\nicefrac{\epsilon_f}{5}$ (i.e. $\epsilon_f$ divided by $2.5$ and $5$), respectively.
This yields an inner approximation of \cref{eq:rawCC_flow} that ensures feasibility of the the AC OPF with desired confidence $1-\epsilon_f$ and the conservatism of the approximation can be tuned by adapting the divisor ($2.5$ and $5$), \cite{dvorkin2019chance}.
Note that the two-sided chance constraints in \cref{eq:rawCC_pG_i,eq:rawCC_qG_i,eq:rawCC_v_i,eq:rawCC_flow} are expressed as one-sided chance constraints in \cref{det_accc:delta_p_plus_i}--\cref{det_accc:nu_ij_f} since simultaneous violations of both the upper and lower capacity or voltage limits are physically  impossible.
Auxiliary variables $\rho_i^v$, $\rho_{ij}^{f^p}$, $\rho_{ij}^{f^q}$ and constraints \cref{det_accc:nu_i_q,det_accc:nu_i_v,det_accc:nu_ij_f} have been introduced to simplify subsequent derivations.
As a result, \cref{mod:main_acccopf} includes convex quadratic objective and second-order conic constraints. Although it  can be reformulated into a  convex conic program to gain computational tractability, \cite{ben2001lectures}, the form  in \cref{mod:main_acccopf} allows for a clear presentation below.

\section{Risk-Aware Pricing}

The EQV-CC endogenously trades off the expected operating point $(p_{G}, q_{G}, v, \theta, \gamma, \alpha)$ and the risk of system limit violations defined by the choice of parameters $z_{\epsilon_g}, z_{\epsilon_q}, z_{\epsilon_v}, z_{\nicefrac{\epsilon_f}{2.5}},z_{\nicefrac{\epsilon_f}{5}}$.
Since the EQV-CC is a convex program, we can use its dual form to compute the marginal prices for active and reactive power, and balancing reserve that internalize this trade-off. 

\subsection{Prices with Chance Constraints on Generation}

First, we consider a modification of the EQV-CC given as:
\allowdisplaybreaks
\begin{subequations}
\begin{align}
  & \text{GEN-CC}: &&\min_{\substack{p_{G}, q_{G} \\ v, \alpha, \theta}} \sum_{i\in \set{N}} c_i (p_{G,i}) + \sum_{i\in \set{N}} \frac{\alpha_i^2}{2b_i} S^2 \\
  & \text{s.t.} && \text{\cref{det_accc:lambda,det_accc:beta,det_accc:delta_p_plus_i,det_accc:delta_p_minus_i,det_accc:chi}} \nonumber \\
  & (\delta_i^{q,+}, \delta_i^{q,-}): && q_{G,i}^{min} \leq q_{G,i} \leq q_{G,i}^{max} \label{gen_cc:delta_i_q_pm} \\
  & (\mu_i^-, \mu_i^+): && v_i^{min} \leq v_i \leq v_i^{max} \label{gen_cc:mu_i_pm} \\
  & (\eta_{ij}): && (f_{ij}^p)^2 + (f_{ij}^q)^2 \leq (s_{ij}^{max})^2, \label{gen_cc:eta_ij}
\end{align}%
\label{mod:gen_cc}%
\end{subequations}%
\allowdisplaybreaks%
where, relative to the EQV-CC in \cref{mod:main_acccopf}, chance constraints are only enforced on active power generation limits and reactive power, voltage and power flow constraints are enforced deterministically by \cref{gen_cc:delta_i_q_pm,gen_cc:mu_i_pm,gen_cc:eta_ij}.
In other words, the GEN-CC determines the optimal balancing participation of each generator and, thus, the optimal amount and allocation of committed reserve given by $\alpha_i z_{\epsilon_g}S$.
Therefore, the GEN-CC replicates a traditional deterministic OPF that allocates the reserve requirement ($\sum_{i\in\set{G}}\alpha_i z_{\epsilon_g}S = z_{\epsilon_g}S$) among individual generators, see \cite{dvorkin2019chance}.

Using the GEN-CC, we compute the following prices:

\begin{prop}
\label{prop:gen_cc_price_decomp}
Consider the GEN-CC in \cref{mod:gen_cc}. Let $\lambda_i^p$, $\lambda_i^q$ be dual multipliers of the nodal active and reactive power balance at node $i$ in \cref{det_accc:lambda}.
Then $\lambda_i^p$ and $\lambda_i^q$ are given as: 
% functions of the marginal cost and scarcity factors ($\delta_i^{p,+}$, $\delta_i^{p,-}$, $\delta_i^{q,+}$, $\delta_i^{q,-}$) of the generator at node $i$:
\begin{align}
  & \lambda_i^p = \frac{p_{G,i} + a_i}{b_i} + \delta_i^{p,-} - \delta_i^{p,+} \label{eq:lambda_i_p_decomp1}\\
  & \lambda_i^q = \delta_i^{q,-} - \delta_i^{q,+}. \label{eq:lambda_i_q_decomp1}
\end{align}

\end{prop}
% \begin{proof}
\noindent
  \textit{Proof.} The first order optimality conditions of \cref{mod:gen_cc} for $p_{G,i}$, $q_{G,i}$, $\alpha_i$, $f_{ij}^p$, $f_{ij}^q$ are:
  \allowdisplaybreaks
  \begin{subequations}
  \begin{align}
    & (p_{G,i}): &&
      \lambda_i^p + (\delta_i^{p,+} -\delta_i^{p,-}) = \frac{p_{G,i} + a_i}{b_i } && i \in \set{G} \label{gencc_kkt:pG_i}\\
    & (q_{G,i}): &&
      \lambda_i^q + (\delta_i^{q,+} -\delta_i^{q,-}) = 0 
      && i \in \set{G} \label{gencc_kkt:qG_i}\\
    & (\alpha_{i}): && 
       z_{\epsilon_p} S (\delta_i^{p,+} + \delta_i^{p,-}) + \chi = \frac{\alpha_i}{b_i}S^2 \hspace{-3cm}&& i \in \set{G} \label{gencc_kkt:alpha_i} \\ 
    & (f_{ij}^p): && 2f_{ij}^p \eta_{ij} + \beta_{ij}^{f^p} = 0 
    \hspace{-3cm}&& ij \in \set{L} \label{gencc_kkt:fp_ij} \\
    & (f_{ij}^q): && 2f_{ij}^q \eta_{ij} + \beta_{ij}^{f^q} = 0 
    \hspace{-3cm}&& ij \in \set{L}. \label{gencc_kkt:fq_ij} 
  \end{align}
  \end{subequations}%
  \allowdisplaybreaks[0]%
  Eqs.~\cref{eq:lambda_i_p_decomp1,eq:lambda_i_q_decomp1} follow directly from \cref{gencc_kkt:pG_i,gencc_kkt:qG_i}. 
  \hspace*{\fill} $\square$
  \vspace{0.3\baselineskip}
% \end{proof}%

Dual multiplier $\lambda_i^p$ of the active power balance, itemized in \cref{eq:lambda_i_p_decomp1}, is interpreted as the real power LMP at node $i$ and a function of production cost coefficients $a_i, b_i$ and scarcity rent $\delta_i^{p,+}$, $\delta_i^{p,-}$ related to active generation limits. 
Dual multiplier $\lambda_i^q$ of the reactive power balance, itemized in \cref{eq:lambda_i_q_decomp1}, is interpreted as the reactive power LMP given by scarcity rent $\delta_i^{q,+}$, $\delta_i^{q,-}$ related to reactive generation limits.
Although there is no explicit production cost for reactive power in \cref{det_accc:objecive}, providing reactive power can have a non-zero value, if at least one reactive power limit is binding. 
Further, Proposition~\ref{prop:gen_cc_price_decomp} shows that both $\lambda_i^p$ and $\lambda_i^q$ in \cref{eq:lambda_i_p_decomp1,eq:lambda_i_q_decomp1} do not explicitly depend on uncertainty and risk parameters. 

In contrast, the price for balancing reserve explicitly depends on the uncertainty and set risk levels:

\begin{prop}
\label{prop:balancing_price_gencc}
Consider the GEN-CC in \cref{mod:gen_cc}. 
Let $\chi$ be the dual multiplier of the balancing adequacy constraint in \cref{det_accc:chi}. 
Then $\chi$ is given as:
\begin{equation}
\begin{split}
  \chi = 
    & \frac{1}{\sum_{i\in\set{G}} b_i} \Big(S^2 + z_{\epsilon_p} S\sum_{i\in\set{G}}b_i(\delta_i^{p,+} + \delta_i^{p,-})\Big). \label{eq:gencc_balprice}
\end{split}
\end{equation}
\end{prop}
% \begin{proof}
\noindent\textit{Proof. }
  Using \cref{det_accc:chi} to eliminate $\alpha_i$ in \cref{gencc_kkt:alpha_i} yields \cref{eq:gencc_balprice}. 
  \hspace*{\fill} $\square$ 
  \vspace{0.3\baselineskip}
% \end{proof}

Dual multiplier $\chi$ of \cref{det_accc:chi} is interpreted as the price for balancing reserve, because it enforces sufficiency of the system-wide reserve.
As per \cref{eq:gencc_balprice}, $\chi$ is an explicit function of the uncertainty $S^2 = e^{\!\top}\Sigma e$ and risk parameter $z_{\epsilon_g}$. 
Notably, the balancing reserve price is always non-zero, if there is uncertainty in the system (i.e. $S > 0$), even if all constraints \cref{det_accc:delta_p_plus_i,det_accc:delta_p_minus_i} are inactive, i.e. $\delta_i^{p,+} = \delta_i^{p,-} 0, \forall i \in \set{G}$.
In this case, $\chi$ is independent of the risk parameters and is determined by the total uncertainty $S^2$ weighted by the total marginal generator cost $\sum_{i\in\set{G}}b_i$ of all generators, $i \in \set{G}$, including those generators that do not provide any balancing reserve, i.e. $\alpha_i =0$.

\subsection{Prices with Complete Chance Constraints}
\label{ssec:complete_chance_constraints}

We now consider the complete EQV-CC in \cref{mod:main_acccopf}, i.e. including chance constraints on reactive power generation, voltages and flows, and prove the following proposition:

\begin{prop}
\label{prop:eqvcc_prices}
Consider the EQV-CC in \cref{mod:main_acccopf}.
Let $\lambda_i^p$, $\lambda_i^q$ be dual multipliers of the nodal active and reactive power balances at node $i$ as in \cref{det_accc:lambda}.
Further, let $\chi$ be the dual multiplier of the balancing adequacy constraint in \cref{det_accc:chi}.
Then (i) $\lambda_i^p$ and $\lambda_i^q$ are given as \cref{eq:lambda_i_p_decomp1,eq:lambda_i_q_decomp1} and (ii) $\chi$ is given as:
\begin{equation}
\begin{split}
\chi\!\! =\!\! 
    \frac{1}{\sum_{i\in\set{G}}\!b_i}\!\!
    \overbrace{\Big(\!\!S^2\!\!+\!\!z_\epsilon S\!\sum_{i\in\set{G}}b_i(\delta_i^+\!\!+\!\delta_i^-\!)}^{\text{Influenced by generator decisions}} 
    +\!\!\!\overbrace{\sum_{i\in\set{G}} b_i (y^q_i\!\!+\!y^v_i\!+\!y^{f^p}\!\!\!\!\!+\!y^{f^q})\!\!\Big)}^{\text{Influenced by system decisions}}, \label{eq:eqvcc_balprice}
\end{split}
\end{equation}
where:
\begin{align}
y^q_i &= 
      z_{\epsilon_q}\!\sum_{j\in\set{G}} [R_j^q]_i \delta_j^{q} \frac{(R_j^q\!+\!\! X_j^q\diag(\gamma))\Sigma e\!-\!R_j^q \alpha S^2}{\sigma_{q_{G,j}}(\alpha, \gamma)} \label{eq:y_q_i} \\
 y^v_i &= 
    z_{\epsilon_v} \sum_{j\in\set{N}} [R_j^v]_i \mu_j \frac{(R_j^v\!+\!\! X_j^v\diag(\gamma))\Sigma e\!-\!R_j^v \alpha S^2}{\sigma_{v_{j}}(\alpha, \gamma)} \label{eq:y_v_i}\\
 y^\diamond_i &= 
    2 \sum_{jk\in\set{L}} [R_{jk}^\diamond]_i \zeta_{ij}^{\diamond}\frac{(R_{jk}^\diamond\!+\!\! X_{jk}^\diamond\diag(\gamma))\Sigma e\!-\!R_{jk}^\diamond \alpha S^2}{\sigma_{\diamond_{jk}}(\alpha, \gamma)}, \label{eq:y_f_ij}
\end{align}
where $\diamond = f^p, f^q$ and $\delta_j^{q} = \delta_j^{q,+} + \delta_j^{q,-}$ , $\mu_j = \mu_j^+ + \mu_j^-$, and $\zeta_{ij}^{\diamond} =  z_{\nicefrac{\epsilon_f}{2.5}}(\xi_{ij}^{\diamond,+} + \xi_{ij}^{\diamond,-}) + z_{\nicefrac{\epsilon_f}{5}}\xi_{ij}^{\diamond,0}$.
Terms $\sigma_{q_{G,j}}(\alpha, \gamma), \sigma_{v_{j}}(\alpha, \gamma)$, $\sigma_{f^p_{jk}}(\alpha, \gamma)$, $\sigma_{f^q_{jk}}(\alpha, \gamma)$ denote the standard deviations of reactive power at node $j$, voltage at node $j$, active power flow on line $jk$ and reactive power flow on line $jk$, respectively, and $[\cdot]_i$ denotes the $i$-th element of a vector.
\end{prop}
\noindent\textit{Proof. }
The first order optimality conditions of \cref{mod:main_acccopf} for $p_{G,i}$, $q_{G,i}$, $\alpha_i$, $f_{ij}^p$, $f_{ij}^q$ and auxiliary variables 
are:
\allowdisplaybreaks
\begin{subequations}
\begin{align}
 & && \text{\cref{gencc_kkt:pG_i,gencc_kkt:qG_i,gencc_kkt:rho_i_q,gencc_kkt:tq_i}} \nonumber \\
& (\alpha_{i}): &&
     \chi + z_{\epsilon_p} S (\delta_i^{p,+}\!\!+\delta_i^{p,-}) + \!\!\sum_{j\in\set{G}}\nu_j^q [R_j^q]_i +\!\!\sum_{j\in\set{N}}\nu_j^v [R_j^v]_i \hspace{-3cm}&& \nonumber \\
    & && \quad  +\!\!\sum_{jk\in\set{L}}\nu_{jk}^{f^p} [R_{jk}^{f^p}]_i +\!\!\sum_{jk\in\set{L}}\nu_{jk}^{f^q} [R_{jk}^{f^q}]_i = \frac{\alpha_i}{b_i}S^2 \hspace{-2cm}&& \nonumber \\
    & && && i \in \set{G} \label{eqvcc_kkt:alpha_i}\\
& (t_{i}^q): && z_{\epsilon_p}(\delta_i^{q,+} + \delta_i^{q,-}) - \zeta_i^q = 0 
    \hspace{-3cm}&& i \in \set{G} \label{gencc_kkt:tq_i} \\
& (\rho_i^q): &&
      \zeta_i^q \frac{(R_i^q - \rho_i^q e^{\!\top} + X_i^q \diag(\gamma)) \Sigma e}{\soc{(R_i^q - \rho_i^q e^{\!\top} + X_i^q \diag(\gamma)) \Sigmart}} - \nu_i^q = 0 \hspace{-3cm}&&\nonumber \\
      & && &&i \in \set{G} \label{gencc_kkt:rho_i_q} \\ 
& (\rho_i^v): &&
      \zeta_i^v \frac{(R_i^v - \rho_i^v e^{\!\top} + X_i^v \diag(\gamma)) \Sigma e}{\soc{(R_i^v - \rho_i^v e^{\!\top} + X_i^v \diag(\gamma)) \Sigmart}} - \nu_i^v = 0 \hspace{-3cm}&& \nonumber \\
      & && && i \in \set{N} \label{eqvcc_kkt:rho_i_v} \\ 
& (t_i^v): && z_{\epsilon_v}(\mu_i^+ + \mu_i^-) - \zeta_i^v = 0 && i \in \set{N} \label{eqvcc_kkt:t_i_v}\\
& (f_{ij}^p): && 
  \beta_{ij}^{f^p} - \xi_{ij}^{f^p,+} + \xi_{ij}^{p,-} = 0 && ij \in \set{L} \label{eqvcc_kkt:f_ij_p}\\
& (f_{ij}^q): && 
  \beta_{ij}^q - \xi_{ij}^{f^q,+} + \xi_{ij}^{f^q,-} = 0 && ij \in \set{L} \label{eqvcc_kkt:f_ij_q}\\  
& (\rho_{ij}^\diamond): &&
      \zeta_i^v \frac{(R_{ij}^\diamond - \rho_{ij}^\diamond e^{\!\top} + X_{ij}^\diamond \diag(\gamma)) \Sigma e}{\soc{(R_{ij}^\diamond - \rho_i^v e^{\!\top} + X_{ij}^\diamond \diag(\gamma)) \Sigmart}} - \nu_{ij}^\diamond = 0 \hspace{-3cm}&&\nonumber \\
      & && &&\hspace{-2cm} ij \in \set{L}, \diamond = f^p, f^q \label{eqvcc_kkt:rho_ij_f} \\ 
& (a_{ij}^{\diamond}): && 
  2\eta_{ij}a_{ij}^{\diamond} - (\xi_{ij}^{\diamond,+} + \xi_{ij}^{\diamond,-}) - \xi_{ij}^{\diamond,0} = 0 && \nonumber \\
  & && && \hspace{-2cm} ij \in \set{L}, \diamond = f^p, f^q \label{eqvcc_kkt:a_ij_p} \\
& (t_{ij}^{\diamond}): && 
  z_{\nicefrac{\epsilon_f}{2.5}}(\xi_{ij}^{\diamond,+} + \xi_{ij}^{\diamond,-}) + z_{\nicefrac{\epsilon_f}{5}}\xi_{ij}^{\diamond,0} - \zeta_{ij}^{\diamond} = 0 \hspace{-2cm}&& \nonumber \\
  & && && \hspace{-2cm} ij \in \set{L}, \diamond = f^p, f^q \label{eqvcc_kkt:t_ij_fp} 
\allowdisplaybreaks[0]%
\end{align}%
\end{subequations}%
The result (i) follows directly from the proof of Proposition~\ref{prop:gen_cc_price_decomp}.  
The result (ii) follows from \cref{eqvcc_kkt:alpha_i} by eliminating $\alpha_i$ using \cref{det_accc:chi}.
Note that terms $\nu_i^q$, $\nu_i^v$, $\nu_{ij}^{f^p}$, $\nu_{ij}^{f^q}$ are given by \cref{gencc_kkt:rho_i_q}, \cref{eqvcc_kkt:rho_i_v} and \cref{eqvcc_kkt:rho_ij_f}.
Further, $t_i^q = \sigma_{q_{G,i}}(\alpha, \gamma)$, if $\zeta_i^q > 0$ as per \cref{det_accc:zeta_i_q},  $t_i^v = \sigma_{v_{j}}(\alpha, \gamma)$, if $\zeta_i^v > 0$ as per \cref{det_accc:zeta_i_v} and $t_{ij}^\diamond = \sigma_{\diamond_{jk}}(\alpha, \gamma)$, if $\zeta_{ij}^\diamond > 0$ as per \cref{det_accc:zeta_ij_f} for $\diamond = f^p, f^q$. 
Thus, for any $\zeta_i^q, \zeta_i^v, \zeta_{ij}^{f^p}, \zeta_{ij}^{f^q} = 0$ the dependency on the standard deviation would disappear.
Finally, terms $\zeta_i^q, \zeta_i^v, \zeta_{ij}^{f^p}, \zeta_{ij}^{f^q}$ are given by \cref{eqvcc_kkt:t_i_v,gencc_kkt:tq_i,eqvcc_kkt:t_ij_fp}. 
\hspace*{\fill} $\square$
\vspace{0.3\baselineskip}
% \end{proof}

Similar to the result of Proposition~\ref{prop:gen_cc_price_decomp}, prices $\lambda^p_i$ and $\lambda^q_i$ do not explicitly depend on uncertainty and risk parameters. 
On the other hand, relative to \cref{eq:gencc_balprice}, balancing reserve price $\chi$ depends on additional terms $y_i^q$, $y_i^v$, $y_i^{f^p}$, $y_i^{f^q}$, see \cref{eq:eqvcc_balprice}, that relate the balancing reserve provided by each generator at node~$i$ to the risk of reactive power and voltage limits violation at every node $j\in\set{N}$ and to the risk of power flow violations on every line $jk\in\set{L}$.
This risk awareness is not part of the generator decisions, which are only driven by its own production limits and cost, as indicated in \cref{eq:eqvcc_balprice}.
As a result of this incompleteness, given system-wide balancing price $\chi$, generators may elect for balancing participation factors which are sub-optimal from the system perspective.
This can be overcome either by further completing the market in terms of transmission and voltage prices as proposed in \cite{lipka2016running}, or by augmenting the system-wide balancing price to reflect location-specific constraints, e.g. $\tilde{\chi}_i \coloneqq \chi + y_i^q + y_i^v+y_i^{f^p}+y_i^{f^q}$.

\section{Variance-Aware Pricing}

The risk-aware results of the EQV-CC in \cref{mod:main_acccopf} yield solutions with a high variability (variance) of system state variables, which has been shown to complicate real-time operations, \cite{baghsorkhi2012impact,bienstock2018variance}.
The variances of reactive power generation, voltage magnitudes, and active and reactive flows can directly be computed from the standard deviations related to $t_i^q$, $t_i^v$, $t_{ij}^{f^p}$, $t_{ij}^{f^q}$, respectively. 
We introduce the metric $V(t_i^q, t_i^v, t_i^{f^p}, t_i^{f^q})$ that models a connection between the variances and system cost in the following variance-aware formulation:
\begin{align}
 & \text{VA-CC}: && \min_{\substack{p_{G}, q_{G} \\ v, \alpha, \theta }} \sum_{i\in \set{N}} c_i (p_{G,i}) +\!\!\sum_{i\in \set{N}}\!\frac{\alpha_i^2}{b_i} S^2\!\!+\! V(t_i^q, t_i^v, t_{ij}^{f^p}, t_{ij}^{f^q}) \nonumber \\
&  \text{s.t.} && \text{\eqref{det_accc:lambda}--\eqref{det_accc:nu_ij_f}}. \label{mod:va_acccopf}
\end{align}
Specifically, metric $V(\cdot)$ penalizes the variance of state variables and, thus, 
it can be used to trade-off the overall system variance and  the expected operating cost in the system as discussed in \cite{bienstock2018variance}.
We define metric $V(\cdot)$ as:
\begin{equation}
\begin{split}
  &V(t_i^q, t_i^v, t_{ij}^{f^p}, t_{ij}^{f^q}) = \sum_{i\in\set{G}}(\Psi_i^q(t_i^q)^2) + \sum_{i \in \set{N}} \Psi_i^v(t_i^v)^2 \\
  & \qquad \quad  + \sum_{ij \in \set{L}}( \Psi_{ij}^{f^p}(t_{ij}^{f^p})^2 + \Psi_i^{f^q}(t_{ij}^{f^q})^2),
\end{split}
\end{equation}
where $\Psi_i^q$, $\Psi_i^v$, $\Psi_{ij}^{f^p}$, $\Psi_{ij}^{f^q}$ are variance penalty factors in the units of $[\nicefrac{\$}{\operatorname{MVAr}^2}]$, $[\nicefrac{\$}{\operatorname{V}^2}]$, $[\nicefrac{\$}{\operatorname{MW}^2}]$ and $[\nicefrac{\$}{\operatorname{MVAr}^2}]$, respectively. 
Note that active power standard deviation $t_i^p$ is already controlled by the generation cost and the constraints on $\alpha_i$.

\begin{prop}
\label{prop:va_pricing}
Consider the VA-CC in \cref{mod:va_acccopf}. 
Let $\lambda_i^p$, $\lambda_i^q$ be dual multipliers of the nodal active and reactive power balance at node $i$ as in \cref{det_accc:lambda}.
Further, let $\chi$ be the dual multiplier of the balancing adequacy constraint in \cref{det_accc:chi}.
Then (i) $\lambda_i^p$ and $\lambda_i^q$ are given by \cref{eq:lambda_i_p_decomp1,eq:lambda_i_q_decomp1} and (ii) $\chi$ is given as:
\begin{equation}%
\begin{split}
\chi\!\! =\!\! 
    \frac{1}{\sum_{i\in\set{G}}\!b_i}\!\!
    \Big(\!\!S^2\!\!+\!\!z_\epsilon S\!\sum_{i\in\set{G}}b_i(\delta_i^+\!\!+\!\delta_i^-\!)
    +\!\!\!\sum_{i\in\set{G}} b_i (y^q_i\!\!+\!y^v_i\!+\!y^{f^p}\!\!\!\!\!+\!y^{f^q})\!\!\Big),
\label{eq:vacc_balprice}
\end{split}
\end{equation}
where:
\begin{align}
  y^q_i &= 
    \sum_{j\in\set{G}} [R_j^q]_i \zeta_j^q \frac{(R_j^q\!+\!\! X_j^q\diag(\gamma))\Sigma e\!-\!R_j^q \alpha S^2}{\sigma_{q_{G,j}}(\alpha, \gamma)} \\
  y^v_i &= 
    \sum_{j\in\set{N}} [R_j^v]_i \zeta_j^v \frac{(R_j^v\!+\!\! X_j^v\diag(\gamma))\Sigma e\!-\!R_j^v \alpha S^2}{\sigma_{v_{j}}(\alpha, \gamma)} \\
  y^\diamond_i &= 
    2 \sum_{jk\in\set{L}} [R_{jk}^\diamond]_i \zeta_{ij}^{\diamond}\frac{(R_{jk}^\diamond\!+\!\! X_{jk}^\diamond\diag(\gamma))\Sigma e\!-\!R_{jk}^\diamond \alpha S^2}{\sigma_{\diamond_{jk}}(\alpha, \gamma)} \\
  \zeta_j^q &= z_{\epsilon_q}(\delta_j^{q,+} + \delta_j^{q,-}) - 2 \sigma_{q_{G_j}}(\alpha, \gamma) \Psi_j^q \label{eq:zeta_j_q_direct} \\
  \zeta_j^v &= z_{\epsilon_v}(\mu_j^+ + \mu_j^-) - 2 \sigma_{v_j}(\alpha, \gamma) \Psi_j^v \label{eq:zeta_j_v_direct} \\
  \zeta_{jk}^{\diamond} &= z_{\nicefrac{\epsilon_f}{2.5}}(\xi_{ij}^{\diamond,+} + \xi_{ij}^{\diamond,-}) + z_{\nicefrac{\epsilon_f}{5}}\xi_{ij}^{\diamond,0} - 2 \sigma_{\diamond_{jk}}(\alpha, \gamma) \Psi_j^{\diamond} \label{eq:zeta_jk_fp_direct}
\end{align}
where $\diamond = f^p, f^q$.
\end{prop}
% \begin{proof}
\noindent\textit{Proof. }
The first-order optimality conditions of \cref{mod:va_acccopf} for $p_{G,i}$, $q_{G,i}$, $\alpha_i$, $f_{ij}^p$, $f_{ij}^q$ and auxiliary variables 
are:
\begin{subequations}
\begin{align}
& && \text{\cref{gencc_kkt:pG_i,gencc_kkt:qG_i,gencc_kkt:rho_i_q,eqvcc_kkt:rho_i_v,eqvcc_kkt:rho_ij_f,eqvcc_kkt:f_ij_p,eqvcc_kkt:a_ij_p,eqvcc_kkt:f_ij_q}} \hspace{-3cm}&& \nonumber \\
& (\alpha_{i}): &&
     z_{\epsilon_p} S (\delta_i^{p,+} + \delta_i^{p,-}) + \chi + \sum_{j\in\set{G}}\nu_j^q [R_j^q]_i \hspace{-3cm}&& \nonumber \\
    & &&\hspace{-1cm} + \sum_{j\in\set{N}}\nu_j^v [R_j^v]_i + \sum_{jk\in\set{L}}\nu_{jk}^{\diamond} [R_{jk}^{\diamond}]_i = 
    (\frac{1}{b_i} + 2\Psi_i^p)\alpha_i S^2 \hspace{-2cm}&& \nonumber \\
    & && &&\hspace{-2cm} i \in \set{G}, \diamond = f^p, f^q \label{vacc_kkt:alpha_i}\\
& (t_{i}^q): && z_{\epsilon_p}(\delta_i^{q,+} + \delta_i^{q,-}) - \zeta_i^q = 2 t_i^q \Psi_i^q && i \in \set{G} \label{vacc_kkt:t_i_q} \\
& (t_i^v): && z_{\epsilon_v}(\mu_i^+ + \mu_i^-) - \zeta_i^v = 2 t_i^v \Psi_i^v && i \in \set{N} \label{vacc_kkt:t_i_v} \\
& (t_{ij}^{\diamond}): && 
  z_{\nicefrac{\epsilon_f}{2.5}}(\xi_{ij}^{\diamond,+} + \xi_{ij}^{\diamond,-}) + z_{\nicefrac{\epsilon_f}{5}}\xi_{ij}^{\diamond,0} - \zeta_{ij}^{\diamond} = 2 t_{ij}^{\diamond} \Psi_{ij}^{\diamond} \hspace{-2cm}&& \nonumber \\
  & && && \hspace{-2cm} ij \in \set{L}, \diamond = f^p, f^q \label{vacc_kkt:t_ij_fp} 
\end{align} 
\end{subequations}
The result (i) follows directly from the proof of Proposition~\ref{prop:gen_cc_price_decomp}.
The result (ii) follows from re-arranging \cref{vacc_kkt:alpha_i} using \cref{det_accc:chi} to eliminate $\alpha_i$.
Note that terms $\nu_i^q$, $\nu_i^v$, $\nu_{ij}^{f^p}$, $\nu_{ij}^{f^q}$ are given by \cref{gencc_kkt:rho_i_q}, \cref{eqvcc_kkt:rho_i_v} and \cref{eqvcc_kkt:rho_ij_f} and terms \cref{eq:zeta_j_q_direct,eq:zeta_j_v_direct,eq:zeta_jk_fp_direct} follow from \cref{vacc_kkt:t_i_q,vacc_kkt:t_i_v,vacc_kkt:t_ij_fp}.
Similarly to the proof of Proposition~\ref{prop:eqvcc_prices}, $t_i^v = \sigma_{v_{j}}(\alpha, \gamma)$, if $\zeta_i^v > 0$ as per \cref{det_accc:zeta_i_v}, $t_{ij}^{f^p} = \sigma_{{f^p}_{jk}}(\alpha, \gamma)$, if $\zeta_{ij}^{f^p} > 0$ as per \cref{det_accc:zeta_i_v}, and $t_{ij}^{f^q} = \sigma_{{f^p}_{jk}}(\alpha, \gamma)$, if $\zeta_{ij}^{f^q} > 0$ as per \cref{det_accc:zeta_ij_f}.
\hspace*{\fill} $\square$
\vspace{0.3\baselineskip}
% \end{proof}

Relative to the results of Proposition~\ref{prop:eqvcc_prices}, terms $y_i^q$, $y_i^v$, $y_i^{f^p}$, $y_i^{f^q}$ now include an inherent trade-off between the risk of limit violation and the absolute standard deviations weighted by penalty factors $\Psi_i^p$, $\Psi_i^q$, $\Psi_i^v$, $\Psi_{ij}^{f^p}$, $\Psi_{ij}^{f^q}$, see \cref{eq:zeta_j_q_direct,eq:zeta_j_v_direct,eq:zeta_jk_fp_direct}.
Since dual multipliers $\zeta_j^q, \zeta_j^v, \zeta_{jk}^{f^p}, \zeta_{jk}^{f^q}$ must be non-negative by definition, the scarcity rents of reactive power $\delta_j^{q,+}, \delta_j^{q,-}$, voltage magnitude $\mu_j^+, \mu_j^-$, active power flows $\xi_{ij}^{f^p,+},\xi_{ij}^{p,-},\xi_{ij}^{f^p,0}$ and reactive power flows $\xi_{ij}^{f^q,+},\xi_{ij}^{f^q,-},\xi_{ij}^{f^q,0}$ and risk parameters $z_{\epsilon_g}, z_{\epsilon_v}, z_{\epsilon_{f}}$ set an upper bound to the standard deviations $\sigma_{p_{G,j}},\sigma_{v_{j}},\sigma_{f^p_{jk}},\sigma_{f^q_{jk}}$ weighted by the penalty factors.

\begin{table*}[t]
    \caption{Optimal Solutions of the deterministic, GEN-CC, EQV-CC and VA-CC cases.}
    \label{tab:results_overview}
    \centering
    \setlength\tabcolsep{5pt} % default value: 6pt
    \begin{tabular}{c|l|c|c|c|c c c c c}
    \toprule
    \multirow{2}{*}[-0.3em]{Risk Level} & Model & Det & GEN-CC & EQV-CC & \multicolumn{5}{c}{VA-CC ($\Psi = \Psi_i^p = \Psi_i^q = \Psi_i^v = \Psi_{ij}^{f^p} = \Psi_{ij}^{f^q}, \forall{i}, \forall{ij}$)} \\
    \cmidrule{2-10}
    & $\Psi$ & -- & -- & -- & 0.1 & 1 & 10 & 100 & 1000 \\
    \midrule\midrule
% eps = 0.1
    \multirow{8}{*}[-0.5em]{\rotatebox[origin=c]{90}{$\epsilon_p = \epsilon_q = \epsilon_v = \epsilon_f  = 0.1$}} & Objective [\$] 
    &  91103.22 & 91107.33 & 92237.67 & 92237.74 & 92238.30 & 92243.86 & 92296.91 & 92764.30 \\
    \cmidrule{2-10}
    & Exp. Gen. Cost [\$] & 91103.22 & 91107.33 & 92237.67 &  92237.68 & 92237.68 & 92237.72 & 92239.70 & 92260.83\\
    & $\Delta$ rel. to EQV-CC & 98.770\% & 98.774\% & 100.000\% & 100.000\% & 100.000\% & 100.000\% & 100.002\% & 100.025\% \\ 
    \cmidrule{2-10}
    & $\chi$ [\$] & -- & 8.72 & 28.10 & 28.11 & 28.23 & 29.40 & 40.35 & 125.54 \\
    \cmidrule{2-10}
    % & $\Delta\sum_i \sigma_{p_{G,i}}^2$ [\%]    & -- & --  & 100.0\%  & 99.981\% & 99.717\% & 97.208\% & 81.940\% & 61.163\% \\
    & $\Delta\sum_i \sigma_{q_{G,i}}^2$ [\%]    & -- & --  & 100.0\% & 0.132\% & 0.103\% & 0.090\% & 0.087\% & 0.064\% \\
    & $\Delta\sum_i \sigma_{v_{i}}^2$ [\%]       & -- & --  & 100.0\% & 3.459\% & 1.215\% & 0.349\% & 0.269\% & 0.225\% \\
    & $\Delta\sum_{ij} \sigma_{f^p_{ij}}^2$ [\%] & -- & --  & 100.0\% & 61.071\% & 60.458\% & 60.537\% & 59.798\% & 59.614\% \\
    & $\Delta\sum_{ij} \sigma_{f^q_{ij}}^2$ [\%] & -- & --  & 100.0\% & 55.808\% & 54.793\% & 54.925\% & 54.584\% & 54.313\% \\
    \midrule\midrule
% eps = 0.01
    \multirow{8}{*}[-0.3em]{\rotatebox[origin=c]{90}{$\epsilon_p = \epsilon_q = \epsilon_v = \epsilon_f  = 0.01$}} 
    & Objective [\$] &  91103.22 & 91107.71 & 93744.95 & 93745.01 & 93745.57 & 93751.17 & 93805.19 & 94281.35 \\
    \cmidrule{2-10}
    & Exp. Gen. Cost [\$] & 91103.22 & 91107.71 & 93744.95 & 93744.95 & 93744.94 & 93744.96 & 93747.04 & 93772.27\\
    & $\Delta$ rel. to EQV-CC  & 97.182\% & 97.187\% & 100.000\% & 100.000\% & 100.000\% & 100.000\% & 100.002\% & 100.029\%\\ 
    \cmidrule{2-10}
    & $\chi$ [\$] & -- & 9.74 & 25.93 & 25.94 & 26.03 & 26.95 & 37.47 & 126.42 \\
    \cmidrule{2-10}
    % & $\Delta\sum_i \sigma_{p_{G,i}}^2$ [\%]    & -- & --  & 100.0\% & 99.995\% & 99.920\% & 99.160\% & 86.065\% & 58.819\%  \\
    & $\Delta\sum_i \sigma_{q_{G,i}}^2$ [\%]    & -- & --  & 100.0\% & 0.194\%  & 0.188\%  & 0.187\%  & 0.163\%  & 0.149\%  \\
    & $\Delta\sum_i \sigma_{v_{i}}^2$  [\%]     & -- & --  & 100.0\% & 25.384\% & 4.570\%  & 1.073\%  & 0.752\%  & 0.650\%  \\
    & $\Delta\sum_{ij} \sigma_{f^p_{ij}}^2$ [\%] & -- & --  & 100.0\% & 64.291\% & 64.526\% & 64.404\% & 62.879\% & 62.103\%  \\
    & $\Delta\sum_{ij} \sigma_{f^q_{ij}}^2$ [\%] & -- & --  & 100.0\% & 54.022\% & 54.241\% & 54.193\% & 52.940\% & 52.626\%  \\
    \bottomrule
    \end{tabular}
\end{table*}

\begin{figure*}[t]
    \centering
    \includegraphics[width=0.9\textwidth]{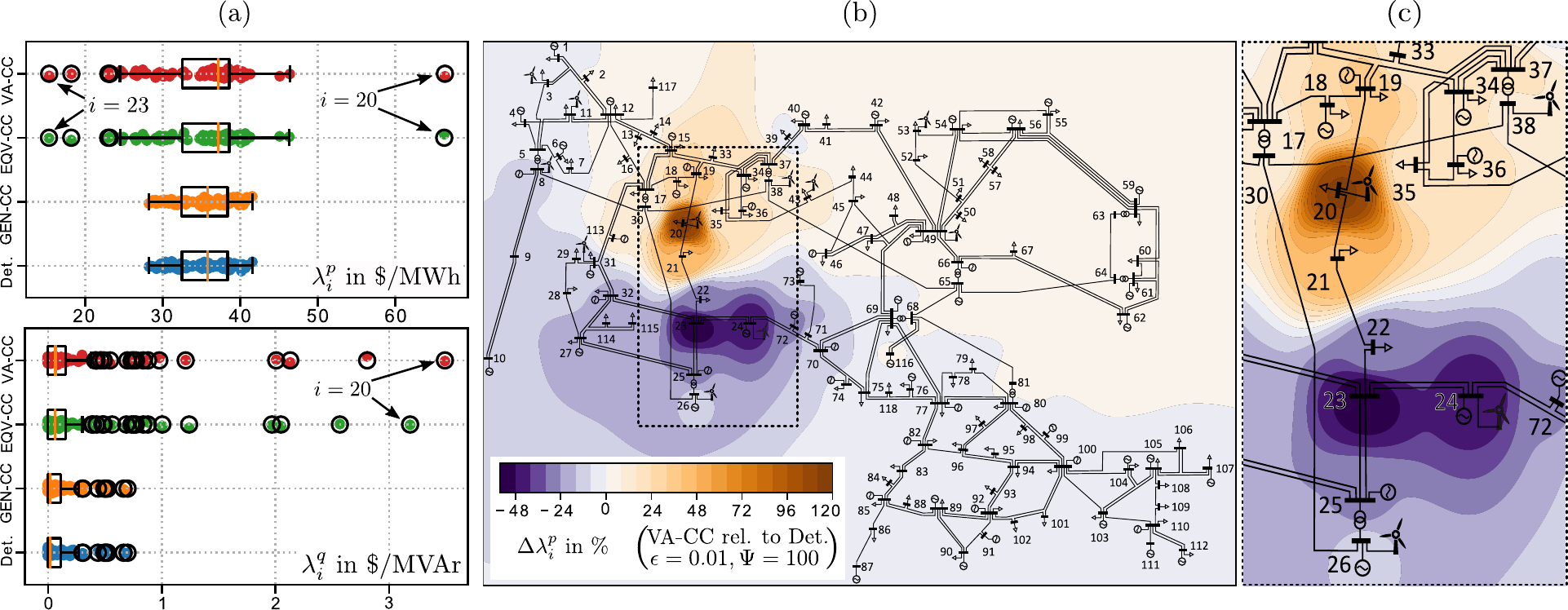}
    \caption{(a) Active and reactive power prices $\lambda_i^p$ and $\lambda_i^q$ for the deterministic, GEN-CC and EQV-CC cases and VA-CC with $\Psi=100$ for risk level $\epsilon = 0.01$. The orange line within the blue box represents the median value, the left and right edges of the box represent the first and third quartiles and the outliers are plotted as circles. 
    (b) Difference of active power prices $\lambda_i^p$ in the VA-CC ($\Psi=100$) relative to the deterministic case (in \%).
    (c) Magnification of the area indicated by the doted rectangle in (b).}
    \label{fig:lambda_mainfig}
\end{figure*}

\section{Case Study}

We conduct numerical experiments using the modified \mbox{118-node} IEEE test system from \cite{dvorkin2019chance}, which includes 11 wind farms with the total forecast power output of \unit[1196]{MW} ($\approx \unit[28.2]{\%}$ of the total active power demand). 
As in \cite{dvorkin2019chance,lubin2016robust}, the wind power forecast error is zero-mean with the standard deviation of $\sigma_{p_{U,i}} = 0.125 p_{U,i}, \forall i \in \set{U}$.
In addition to the GEN-CC, EQV-CC and VA-CC, we solve a deterministic AC OPF (reference) case using the forecast renewable generation and ${\alpha_i = 0, \forall i \in \set{G}}$.
All calculations have been performed for risk levels $\epsilon = 0.1$ and $\epsilon = 0.01$ assuming that $\epsilon_p = \epsilon_q = \epsilon_v = \epsilon_f = \epsilon$.
Additionally, the VA-CC has been computed for various values of $\Psi = \{0.1, 1, 10, 100, 1000\}$ assuming that $\Psi_i^p = \Psi_i^q = \Psi_i^v = \Psi, \forall i \in \set{N}$  and  $\Psi_{ij}^{f^p} = \Psi_{ij}^{f^q} = \Psi, \forall{ij} \in \set{L}$.
All models are implemented in Julia using JuMP \cite{jump} and the  code and input data are reported in \cite{acccopf_pricing_code}. 
The linearization point (see Section~\ref{ssec:linearization_of_pf_equations}) has been obtained as described in \cite{dvorkin2019chance} using the IPOPT solver, \cite{ipopt}, and the chance-constrained models have been solved using the MOSEK solver, \cite{mosek}.

\subsection{Cost and Price Analysis}

Table~\ref{tab:results_overview} compares the results of the deterministic, GEN-CC, EQV-CC and VA-CC cases for different values of $\epsilon$ and $\Psi$.  As expected, the objective value and expected generation cost increase as we introduce additional chance constraints and increase the value of $\Psi$, thus internalizing the cost of re-dispatch to ensure larger security margins and lower variance of state variables. 
Similarly to the results in \cite{bienstock2018variance}, which uses DC power flow assumptions, increasing variance penalty factor $\Psi$ does not significantly raise the expected generation cost.  
This observation suggests that this reduction in state variable variances is achieved by adjustments to those variables which are not limited by binding constraints in the optimal solution. In other words, the variance of variables related to non-binding constraints can be controlled without significantly affecting the optimal values of other variables. Note that the variance of variables related to binding chance constraints is \textit{a priori} controlled  by the violation tolerance  of these constraints.

Also, increasing conservatism of the model increases system-wide balancing reserve price $\chi$ for both values of $\epsilon$. For example, in the GEN-CC, the value of $\chi$ is only driven by chance constraints on power output limits of generators, as per Proposition~\ref{prop:balancing_price_gencc}, while the EQV-CC and VA-CC introduce additional components (e.g. reactive power, voltage and flow variances) to price $\chi$ as per Propositions~\ref{prop:eqvcc_prices} and \ref{prop:va_pricing}. 
Location-specific prices $\lambda_i^p$ and $\lambda_i^q$ for all network nodes are displayed in Fig.~\ref{fig:lambda_mainfig}a), while Figs.~\ref{fig:lambda_mainfig}b)--c) map the relative difference between $\lambda_i^p$ for the VA-CC case with $\Psi=100$ and $\epsilon = 0.01$ and the deterministic case.  
At the majority of nodes, prices $\lambda^p_i$ (indicated by the box-plots in Fig.~\ref{fig:lambda_mainfig}a) remain within \unit[32--38]{\$/MWh}.  Note that unlike $\chi$, which significantly increases for more conservative models, prices for $\lambda^p_i$  and $\lambda^q_i$ do not vary as much as conservatism increases. This corresponds to our findings in Propositions~\ref{prop:gen_cc_price_decomp}--\ref{prop:va_pricing}, which show that active and reactive power prices do not explicitly depend on the uncertainty and risk parameters. 
However, at some nodes, prices $\lambda^p_i$ and $\lambda^q_i$ in the GEN-CC and VA-CC cases exhibit larger deviations, e.g. see $\lambda_i^p$ at nodes 20 and 23, which are also in proximity of wind farms, as shown in Fig.~\ref{fig:lambda_mainfig}c). A resulting high flow variance on the line between nodes 19 and 23 causes price differentiation at nodes \mbox{19, 20, 21} and \mbox{23, 24, 25}. 

\subsection{Analysis of Variance of State Variables}
Table~\ref{tab:results_overview} shows how the aggregated variance of state variables $\sum_i \sigma_{q_{G,i}}^2$, $\sum_i \sigma_{v_{i}}^2$, $\sum_i \sigma_{f^p_{ij}}^2$, $\sum_i \sigma_{f^q_{ij}}^2$ change relative to the EQV-CC case as penalty $\Psi$ increases. Even if $\Psi$ is set to a small value, the variance of state variables reduce significantly, without a large increase in the objective function, expected generation cost, and prices $\lambda^p_i$ and $\lambda^q_i$. Furthermore, as the value of $\epsilon$ increases, the relative reduction in variances of all state variables slightly reduces. The effect of variance penalty $\Psi$ on prices is two-fold. First, it does not affect prices $\lambda_i^p$ and $\lambda_i^q$ relative to the EQV-CC case. Second, system-wide balancing price $\chi$, which internalizes the variance penalties as per Proposition~\ref{prop:va_pricing}, increases with penalty $\Psi$.  

\section{Conclusion}

This paper described an approach to internalize RES stochasticity and risk parameters in electricity prices.
Using SOC duality, these risk- and variance-aware prices are derived from a chance-constrained AC-OPF and are itemized in terms of active and reactive power, voltage support and power flow components. 
We proved that active and reactive power prices do not explicitly depend on uncertainty and risk parameters, while expressions for balancing reserve prices explicitly include these parameters. 
Further, introducing variance penalties on the system state variables has been shown to internalize the trade-off between variance, risk and system cost at a modest increase in the expected operating cost. 
The results have been demonstrated and analyzed on the modified IEEE 118-node testbed.
Future work includes extensions of the proposed market-clearing model to  account for  risk-averse strategies of market participants,  enable risk trading instruments using our preliminary work in \cite{mieth2020risk}, and to account for multi-period trading horizons.

\bibliographystyle{IEEEtran}
\bibliography{literature_local}

\newpage
\numberwithin{equation}{section}
\appendices
\section{}
\label{ax:sensitivity_derivation}

Rewrite \cref{eq:p_i_linear,eq:q_i_linear} in the following form:
\begin{align}
  \begin{bmatrix} p(\om) \\ q(\om) \end{bmatrix} - \begin{bmatrix} \bar{p} \\ \bar{q} \end{bmatrix} = \begin{bmatrix}J^{p,v} & J^{p,\theta} \\ J^{q,v} & J^{q,\theta} \end{bmatrix} \begin{bmatrix} v(\om) \\ \theta(\om) \end{bmatrix} = J \begin{bmatrix} v(\om) \\ \theta(\om) \end{bmatrix}, \label{ax:lin_pb_matrix} 
\end{align}
where the rows of matrices $J^\diamond$ are equal to sensitivity vectors $J^\diamond_i$ for $i\in\set{N}$ and $\diamond=\{(p,v);(p,\theta);(q,v);(q,\theta)\}$.
First, we sort the rows of the terms in \cref{ax:lin_pb_matrix} by node types and introduce superscripts $PQ$, $PV$, $\theta V$ to indicate the node type:
\begin{align}
  \begin{matrix}
  \begin{bmatrix} p^{PQ}(\om) \\ p^{PV}(\om) \\ q^{PQ}(\om) \end{bmatrix} \vspace{2pt} \\
  \begin{bmatrix} p^{\theta V}(\om) \\ q^{PV}(\om) \\ q^{\theta V}(\om) \end{bmatrix} \\
  \end{matrix}
   -  \begin{matrix}
  \begin{bmatrix} \bar{p}^{PQ} \\ \bar{p}^{PV} \\ \bar{q}^{PQ} \end{bmatrix} \vspace{2pt} \\
  \begin{bmatrix} \bar{p}^{\theta V} \\ \bar{q}^{PV} \\ \bar{q}^{\theta V} \end{bmatrix} \\
  \end{matrix}
  = \begin{bmatrix}
  J^A & J^B \\ J^C & J^D 
  \end{bmatrix}
  \begin{matrix}
  \begin{bmatrix} v^{PQ}(\om) \\ \theta^{PQ}(\om) \\ \theta^{PV}(\om) \end{bmatrix} \vspace{2pt} \\
  \begin{bmatrix} v^{PV}(\om) \\ v^{\theta V}(\om) \\ \theta^{\theta V}(\om) \end{bmatrix} \\
  \end{matrix}, \label{ax:matrix_pb_sorted}
\end{align}
where $J^{A-D}$ denote the blocks of re-arranged matrix $J$ from \cref{ax:lin_pb_matrix}.
Quantities $p^{PQ}(\om), p^{PV}(\om), q^{PQ}(\om)$ are explicitly given by the uncertain generation and the respective system responses such that:
\begin{align}
  \begin{bmatrix} p^{PQ}(\om) \\ p^{PV}(\om) \\ q^{PQ}(\om) \end{bmatrix}\!-\!\begin{bmatrix} \bar{p}^{PQ} \\ \bar{p}^{PV} \\ \bar{q}^{PQ} \end{bmatrix}
  =
  \begin{bmatrix} p_{G}^{PQ} \\
          p_{G}^{PV} \\
          q_{G}^{PQ}
  \end{bmatrix} \!+\! 
  \begin{bmatrix} (\om + \alpha\Om)^{PQ} \\
          (\om + \alpha\Om)^{PV} \\
          (\diag(\gamma)\om)^{PQ}
  \end{bmatrix}.
  \label{ax:explicit_pq}
\end{align}
Notably, $p_U$ and $p_D$ are not part of the right-hand side of \cref{ax:explicit_pq} because they are fixed parameters.
Further, $v^{PV}(\om) = v^{PV}$, $v^{\theta V}(\om) = v^{PV}$, and $\theta^{\theta V}(\om) = \theta^{\theta V}$ as discussed in Section~\ref{ssec:system_response}.
We use this relationship and \cref{ax:matrix_pb_sorted,ax:explicit_pq} to compute the reactions of the uncontrolled variables to uncertainty $\om$:
\begin{align}
  \begin{bmatrix} v^{PQ}(\om) \\ \theta^{PQ}(\om) \\ \theta^{PV}(\om) \end{bmatrix} -
  \begin{bmatrix} v^{PQ} \\ \theta^{PQ} \\ \theta^{PV} \end{bmatrix}
  = (J^A)^{-1} \begin{bmatrix} (\om + \alpha\Om)^{PQ} \\
          (\om + \alpha\Om)^{PV} \\
          (\diag(\gamma)\om)^{PQ}
  \end{bmatrix}.
  \label{ax:implicit_vtheta}
\end{align}
Note that although $v^{PQ}, \theta^{PQ}, \theta^{PV}$ implicitly depend on the AC power flow equations, these variables are endogenous to the model and not subject to uncertainty.
Similarly, we get:
\begin{equation}
\begin{split}
  \begin{bmatrix} p^{\theta V}\!(\om) \\ q^{PV}\!(\om) \\ q^{\theta V}\!(\om) \end{bmatrix}\!\!-\!\!\begin{bmatrix} p^{\theta V} \\ q^{PV} \\ q^{\theta V} \end{bmatrix}\!\!-\!\!\begin{bmatrix} \bar{p}^{\theta V} \\ \bar{p}^{PV} \\ \bar{q}^{\theta V} \end{bmatrix} 
\!\!\!=\!\!J^{C}\!(J^A)^{\!-1}\!\!\begin{bmatrix} (\om + \alpha\Om)^{PQ} \\
          (\om + \alpha\Om)^{PV} \\
          (\diag(\gamma)\om)^{PQ} 
          \end{bmatrix}.
\end{split}
 \label{ax:implicit_pq}
\end{equation}
Using \cref{ax:implicit_vtheta}, we immediately obtain \cref{eq:v_i_sensitivity} by separating matrix $(J^A)^{-1}$. 
Similarly, we obtain \cref{eq:qG_i_sensitivity} from separating matrix $J^C(J^A)^{-1}$.
In analogy, \cref{eq:fp_ij_sensitivity,eq:fq_ij_sensitivity} can be obtained by noting that $p_i = \sum_{j:ij\in\set{L}} f_{ij}^p$ and $q_i = \sum_{j:ij\in\set{L}} f_{ij}^q$ and combining the sensitivity factors respectively.

\end{document}